\documentclass[twoside]{ae100prg}
\usepackage{graphicx}
\usepackage[breaklinks]{hyperref}
\usepackage{booktabs}

\def\ben{\begin{equation}}
\def\een{\end{equation}}
\def\half{\frac{1}{2}}
\def\R{{{\mathbb R}}}
\def\SU{{\rm SU}}
\def\SO{{\rm SO}}
\def\ISO{{\rm ISO}}

\begin{document}
\title{Spontaneous breaking of Lorentz symmetry for canonical gravity}

\author{Steffen Gielen}
\address{Perimeter Institute for Theoretical Physics, 31 Caroline St. N., Waterloo, Ontario, N2L 2Y5, Canada 
}
\email{sgielen@perimeterinstitute.ca}

\begin{abstract}
In the Ashtekar-Barbero formulation of canonical general relativity based on an $\SU(2)$ connection, Lorentz covariance is a subtle issue which has been the focus of some debate. Here we present a Lorentz covariant formulation generalising the notion of a foliation of spacetime to a field of {\em local observers} which specify a time direction only locally. This field spontaneously breaks the local $\SO(3,1)$ symmetry down to a subgroup $\SO(3)$; we show that the apparent symmetry breaking to $\SO(3)$ is not in conflict with Lorentz covariance. We give a geometric picture of our construction as {\em Cartan geometrodynamics} and outline further applications of the formalism of local observers, motivating the idea that {\em observer space}, instead of spacetime, should serve as the fundamental arena for gravitational physics.
\end{abstract}

\section{Introduction}
In first order formulations of general relativity one has a notion of local Lorentz invariance, which can be thought of as one way of implementing the equivalence principle \footnote{Linking my talk at this wonderful conference to Einstein's Prague days.}.

It is crucial to understand the fate of this gauge symmetry in attempts to quantise gravity, both theoretically and with regard to a possible phenomenology of quantum gravity (including matter). There are strong experimental constraints on many possible types of violation of Lorentz covariance and any proposed theory of quantum gravity must prove itself consistent with such constraints.

In Hamiltonian formulations, in particular the Ashtekar-Barbero connection formulation \cite{ashtekar,barbero}, the issue of Lorentz covariance has been the focus of some debate, since the Ashtekar-Barbero formulation naturally uses the gauge group $\SU(2)$ or $\SO(3)$\footnote{The covering group $\SU(2)$ is required if one wants to include spinors. We consider pure gravity; the symmetry groups we discuss arise as the isometry groups of real manifolds or the stabilisers of points in them, and can be taken to be real-valued matrix groups. By expressions such as $\SO(3,1)$, we mean the connected component preserving orientation and time orientation.}, instead of the full Lorentz group. The use of this smaller gauge group is connected to the appearance of second-class constraints in previous attempts to maintain full Lorentz covariance. Here we show how to avoid second class constraints and stay Lorentz covariant by introducing a field of {\em local observers}. Details are given in the paper \cite{lorentz}.

\section{Canonical First Order General Relativity}
Starting from the Lorentz covariant Palatini-Holst action for vacuum general relativity without cosmological constant
\ben
S[e,\omega]=\frac{1}{8\pi G}\int\kappa_{abcd}\,e^a\wedge e^b\wedge R^{cd}[\omega]\,,
\label{palholst}
\een
where $\kappa_{abcd}$ is an $\SO(3,1)$-invariant bilinear form on $\mathfrak{so}(3,1)$,
\ben
\kappa_{abcd}=\half\epsilon_{abcd}+\frac{1}{2\gamma}(\eta_{ac}\eta_{bd}-\eta_{ad}\eta_{bc})\,,
\een
one can perform the usual canonical analysis and find that the 18 momenta $\pi_{ab}^i$ conjugate to the spatial components of the connection $\omega^{ab}_i$ are expressible in terms of only 12 tetrad components $e^a_i$. This leads to {\em second class constraints}, which provide an obstacle to quantisation and usually require introducing new variables which are harder to interpret in terms of spacetime geometry.

In Holst's analysis \cite{holst} leading to the well-known Ashtekar-Barbero formulation of canonical gravity, one deals with this issue by explicit symmetry breaking to $\SO(3)$: Imposing `time gauge' $e^0_i=0$ and defining
\ben
A^{ab}=\omega^{ab}+\frac{\gamma}{2}{\epsilon^{ab}}_{cd}\omega^{cd}\,,
\een
only the $\mathfrak{so}(3)$ part of $A$ (the {\em Ashtekar-Barbero connection}) has nonvanishing conjugate momentum, and one avoids second class constraints. However, this comes at the price of losing Lorentz symmetry which is broken explicitly by the gauge choice.

In our formalism we replace time gauge by a condition involving a field of {\em internal observers} $y$ which specifies a time direction locally, and leads to a spontaneous breaking of symmetry from $\SO(3,1)$ to a subgroup $\SO(3)_y$ depending on $y(x)$ at each spacetime point $x$.

\section{General Relativity with Local Observers}
For a given spacetime manifold with metric $g$ or frame field $e$, we define a {\em field of observers} as a unit future-directed timelike vector field $u$. Using the frame field we can map it to a spacetime scalar $y=e(u)$ valued in the velocity hyperboloid ${\rm H}^3=\SO(3,1)/\SO(3)$. But such a field of {\em internal observers} can be defined without specifying the metric, and is hence suitable for a framework in which the metric arises dynamically as a solution to the equations of motion.

Our formalism for {\em generalised canonical gravity} builds on the following variables:
\begin{itemize}
\item a field of internal observers $y$, valued in ${\rm H}^3\subseteq \R^{3,1}$, thought of as giving a local notion of time direction,
\item a nowhere-vanishing 1-form $\hat{u}$, thought of as non-dynamical and generalising the normal to a foliation (if $\hat{u}\wedge d\hat{u}=0$, $\hat{u}$ is of the form $\hat{u}=N\,dt$) -- one can always reduce to the case of a foliation by choosing an appropriate $\hat{u}$,
\item an $\R^3_y$-valued `triad' 1-form $E$, where $\R^3_y$ is the subspace of $\R^{3,1}$ orthogonal to $y$ (this generalises time gauge).
\end{itemize}
The spacetime coframe field is then simply given by
\ben
e=E+\hat{u}\,y
\label{one}
\een
analogous to how one reconstructs the spacetime metric in the ADM formulation using lapse and shift. As is usual in first order gravity, we must require $e$ to be nondegenerate. The field of internal observers $y$ defines a field of spacetime observers by $y=e(u)$, and one finds that $E(u)=0$ so that $E$ is actually {\em spatial}.

Similarly, we define spatial and temporal parts of the spin connection,
\ben
\omega = \Omega + \hat{u}\,\Xi
\label{two}
\een
Substituting (\ref{one}) and (\ref{two}) into the Palatini-Holst action (\ref{palholst}) gives us a generalised Hamiltonian formulation of vacuum general relativity in terms of an action depending on $y,E,\Omega$ and $\Xi$ that we give in \cite{lorentz}. Up to this stage everything is Lorentz covariant -- we have just changed variables in the action.

The r\^{o}le of the field of internal observers $y$ is to give us a local embedding of $\SO(3)$ into $\SO(3,1)$. The embedding can be freely changed by applying a Lorentz transformation $y\mapsto y'=\Lambda\,y$; allowing those Lorentz transformations instead of thinking of $y$ as fixed restores Lorentz covariance.

The spatial connection $\Omega$ can be projected to its $\mathfrak{so}(3)_y$ part ${\bf \Omega}$. Then under a local Lorentz transformation
\ben
{\bf \Omega} \mapsto {\bf \Omega}' = \Lambda^{-1}\,{\bf \Omega}\,\Lambda+\pi_{y'}(\Lambda^{-1}\,d^{\perp}\Lambda)\,,
\een
where $\pi_{y'}$ is a projector onto $\mathfrak{so}(3)_{y'}$ and $d^{\perp}=d-\hat{u}\wedge\pounds_u$ is a spatial exterior derivative. Therefore, if one only applies $\SO(3)_y$ transformations which leave $y$ invariant, ${\bf \Omega}$ transforms as an $\SO(3)_y$ connection, while if one allows for transformations that rotate the local internal observer $y$ to $y'$, the transformed connection ${\bf\Omega}'$ is in $\mathfrak{so}(3)_{y'}$. This is as it should be.

To understand the dynamical structure of this formalism, we focus on the term in the action that determines the symplectic structure in Hamiltonian general relativity,
\ben
S=\frac{1}{8\pi G}\int\kappa_{abcd}\hat{u}\wedge E^a\wedge E^b\wedge \pounds_u\Omega^{cd}+\ldots
\een
Since $E\wedge E$ is valued only in $\mathfrak{so}(3)_y$, only half of the components of $\Omega$ have nonvanishing conjugate momentum. The number of independent components of $E$ matches the number of conjugate momenta, and no second-class constraints arise -- but we did not find it necessary to impose any gauge fixing such as the time gauge employed in Holst's analysis.

One can make the splitting of $\mathfrak{so}(3,1)$ into a rotational subalgebra $\mathfrak{so}(3)_y$ and a complement $\mathfrak{p}_y$ explicit by choosing local bases $J^{ab}_I$ and $B^{ab}_I$ (depending on $y$). Then 
\ben
A^I:={\bf \Omega}^I+\gamma K^I\,,
\label{ashbarb}
\een
is conjugate to $(E\wedge E)^I$, where ${\bf \Omega}$ and $K$ are the $\mathfrak{so}(3)_y$ and $\mathfrak{p}_y$ parts of $\Omega$. (\ref{ashbarb}) is the Ashtekar-Barbero connection, and our formalism is dynamically equivalent to the Ashtekar-Barbero formulation: It has the same phase space variables, subject to the same constraints that define the dynamics. In the form (\ref{ashbarb}) manifest Lorentz covariance is lost; it can be recovered by viewing $\mathfrak{so}(3)_y$ and $\mathfrak{p}_y$ not as fixed (isomorphic) representations of $\SO(3)$, but as subspaces of $\mathfrak{so}(3,1)$ specified by the field $y$.

\section{Cartan Geometrodynamics}
Situations of spontaneous symmetry breaking in gravitational theories are geometrically best understood in terms of {\em Cartan geometry} \cite{derekproc}. A well-known example is the MacDowell-Mansouri formulation \cite{macdo} of gravity with cosmological constant (we take $\Lambda>0$ but $\Lambda<0$ is analogous) in terms of the $\SO(4,1)$ invariant action
\ben
S_{\rm MM} = -\frac{3}{32\pi G\Lambda}\int\epsilon_{abcde}\left(F^{ab}\wedge F^{cd}\right)y^e\,,
\label{macdo}
\een
where $F$ is the curvature of an $\SO(4,1)$ connection $A$. The field $y$ takes values in de Sitter spacetime $\SO(4,1)/\SO(3,1)\subseteq \R^{4,1}$; it breaks the symmetry at each point in spacetime to the subgroup $\SO(3,1)_y$ leaving $y$ invariant. Fixing $y=(0,0,0,0,1)$ in the action breaks the symmetry explicitly.

The Lie algebra $\mathfrak{so}(4,1)$ splits into a subalgebra $\mathfrak{so}(3,1)_y$ and a complement $\mathfrak{t}_y$; identifying the $\mathfrak{so}(3,1)_y$ part of $A$ with the spin connection $\omega$ and the $\mathfrak{t}_y$ part with a coframe $e$,
\ben
A = \left(\begin{array}{c c}
 \omega & \sqrt{\frac{\Lambda}{3}}e
\\ -\sqrt{\frac{\Lambda}{3}}e & 0 
\end{array}\right)\,,
\label{cartanco}
\een
the action (\ref{macdo}) reduces to the Einstein-Hilbert-Palatini action with a cosmological term.

Cartan geometry is about infinitesimally approximating the geometry of a curved manifold by a homogeneous spacetime $G/H$ (in this case de Sitter spacetime) which generalises the tangent space $\R^{p,q}$ used in (pseudo-)Riemannian geometry. The Cartan connection $A$ relates the model spacetimes tangent to different points of the manifold -- for a model spacetime of non-zero curvature, $A$ is flat if the manifold is (locally) isomorphic to the model spacetime. This naturally introduces a cosmological constant into gravity, given by the curvature scale of the model spacetime.

Our reformulation of the Ashtekar-Barbero formalism for canonical gravity is best interpreted as describing the geometry of {\em space} as {\em Cartan geometrodynamics}: The $\mathfrak{so}(3)_y$ connection ${\bf \Omega}$ (or, alternatively, the Ashtekar-Barbero connection) and the triad $E$ can be assembled into a Cartan connection
\ben
{\bf A} = \left(\begin{array}{c c}
 {\bf \Omega} & \frac{1}{l}E
\\ 0 & 0 
\end{array}\right)\,,
\label{cartangeo}
\een
taking values in the Lie algebra of the Euclidean group $\mathfrak{iso}(3)$ if we consider a vanishing cosmological constant ($l$ is an (unspecified) length scale put in for dimensional reasons). The appearance of the group $\ISO(3)$ is understood as follows: Spacetime is infinitesimally modelled on Minkowski spacetime, with isometry group $\ISO(3,1)$. At a given point in spacetime, picking an observer in the model Minkowski spacetime gives a notion of `space' in the model spacetime as the maximal totally geodesic hypersurface orthogonal to this observer -- in the construction above, we referred to this as the subspace $\R^3_y$ orthogonal to an observer $y$. This breaks the symmetry to $\ISO(3)$, the isometry group of $\R^3_y$. Picking a point in $\R^3_y$ tangent to the spacetime point then breaks the symmetry further to $\SO(3)$, giving the splitting (\ref{cartangeo}). For a more detailed discussion of the geometry behind Cartan geometrodynamics we refer to \cite{obssp}.

\section{Summary and Outlook}
We have given a reformulation of canonical general relativity in first order form which uses local observers that define a local notion of time. These give an embedding of the rotational subgroup $\SO(3)$ into the Lorentz group that allows to reconstruct Lorentz covariance from the $\SO(3)$ Ashtekar-Barbero formulation of canonical gravity. The geometry behind our constructions is best understood in terms of {\em Cartan geometrodynamics}. Since this formulation requires only a local choice of time direction not necessarily related to a foliation of spacetime, it links the canonical and covariant formulations of general relativity \cite{cancov}.

It would be important to understand the coupling of matter -- which would be necessary to investigate the possibility of physically observable Lorentz violation -- and the role of the field of internal observers there. So far they have been treated like lapse and shift, as Lagrange multipliers. Making the observer field dynamical could relate our framework to models with dynamical reference frames, such as Brown-Kucha\v{r} dust \cite{brownkuch}.

Similar constructions could also be useful in approaches to quantum gravity where local Lorentz covariance is not manifest, such as Ho\v{r}ava-Lifshitz gravity, shape dynamics or causal dynamical triangulations.

Taking the idea of local observers one step further, it is natural to consider the space of all possible choices of local observer -- {\em observer space}. In general relativity, this is the direct product of spacetime with the local velocity space ${\rm H}^3$ of normalised future-directed timelike vectors, but we consider it as a seven-dimensional manifold in its own right and study its geometry, both in general relativity and in more general settings. This is the viewpoint adopted in the work \cite{obssp}, where we show how the Cartan connection $A$ specified by a frame field $e$ and a spin connection $\omega$ as in (\ref{cartanco}) gives a Cartan geometry on observer space, with model space $\SO(4,1)/\SO(3)$, the space of all observers in de Sitter spacetime. Conversely, we investigate integrability conditions that allow the reconstruction of an invariant spacetime starting from an observer space Cartan geometry (i.e. a general Cartan geometry modelled on $\SO(4,1)/\SO(3)$); intuitively, such a reconstruction is possible if the connection is flat in the `velocity' directions of observer space.

Different approaches to quantum gravity and quantum-gravity phenomenology incorporate the idea that spacetime geometry is an observer-dependent (or `momentum-dependent'), relative concept. From the perspective of observer space, such ideas correspond to observer space Cartan connections that are not flat in velocity directions, so that no invariant spacetime can be reconstructed.

One example is the proposal of relative locality \cite{relloc} which suggests that `spacetime' and hence the notion of locality are observer-dependent, but there is an invariant momentum space shared by all observers. In \cite{obssp} we find that the framework of relative locality corresponds to an observer space connection that is flat in `spacetime', not `velocity' directions. For a general observer space geometry, both `spacetime' and `velocity space' are only defined relative to an observer. 

It will be interesting to see whether other ideas, such as that of an `effective metric' $\langle g_{\mu\nu}\rangle_k$ (depending on a momentum scale $k$) that appears in the asymptotic safety scenario for quantum gravity \cite{asympt}, can be discussed in the framework of observer space geometry.

{\bf Acknowledgements.} -- I would like to thank Derek Wise for collaboration on the papers \cite{obssp,cancov,lorentz} that discuss the ideas presented in this proceedings contribution in detail and for comments on the manuscript. Research at Perimeter Institute is supported by the Government of Canada through Industry Canada and by the Province of Ontario through the Ministry of Research \& Innovation.

\end{document}